\documentclass[12pt]{article}
\newcommand{\RRR}{$I\hspace{-0.25em}R^3$}

\newcommand{\ket}[1]{\vert#1\rangle}

\newcommand\sandwich[3]{\langle#1\vert#2\vert#3\rangle}
\newcommand{\be}{\begin{equation}}
\newcommand{\ee}{\end{equation}}
\newcommand{\bea}{\begin{eqnarray}}
\newcommand{\eea}{\end{eqnarray}}
\newcommand{\bq}{\begin{quote}}
\newcommand{\eq}{\end{quote}}

\newcommand{\cS}{{\cal S}}

\newcommand{\cP}{{\cal P}}
\newcommand{\cO}{{\cal O}}
\newcommand{\cQ}{{\cal Q}}

\begin{document}
\setlength{\baselineskip}{16.5pt}
\title{THIS ELUSIVE OBJECTIVE EXISTENCE}
\author{Ulrich Mohrhoff\\
Sri Aurobindo International Centre of Education\\
Pondicherry 605002 India\\
\normalsize\tt ujm@auromail.net}
\date{}
\maketitle
\begin{abstract}
\noindent Zurek's existential interpretation of quantum mechanics suffers from three 
classical prejudices, including the belief that space and time are intrinsically and infinitely 
differentiated. They compel him to relativize the concept of objective existence in two ways. 
The elimination of these prejudices makes it possible to recognize the quantum formalism's 
ontological implications---the relative and contingent reality of spatiotemporal distinctions 
and the extrinsic and finite spatiotemporal differentiation of the physical world---which in 
turn makes it possible to arrive at an unqualified objective existence. Contrary to a 
widespread misconception, viewing the quantum formalism as being fundamentally a 
probability algorithm does not imply that quantum mechanics is concerned with states of 
knowledge rather than states of Nature. On the contrary, it makes possible a complete and 
strongly objective description of the physical world that requires no reference to observers. 
What objectively exists, in a sense that requires no qualification, is the trajectories of 
macroscopic objects, whose fuzziness is empirically irrelevant, the properties and values of 
whose possession these trajectories provide indelible records, and the fuzzy and temporally 
undifferentiated states of affairs that obtain between measurements and are described by 
counterfactual probability assignments.

\vspace{6pt}\noindent {\it Keywords\/}: interpretation of quantum mechanics; 
macroscopic objects; objective existence; pointer states; quantum states; space; time.
\setlength{\baselineskip}{14pt}
\end{abstract}

\section{\large Introduction}
Detailed investigations of a large class of specific models, carried out over the past two 
decades, have demonstrated that the reduced density operators associated with sufficiently 
large and/or massive systems, obtained by partial tracing over realistic environments, 
become very nearly diagonal with respect to a privileged basis in very short times, and that 
they stay that way for very long times. In such systems environment-induced {\it 
decoherence\/} leads to {\it einselection\/} (environment-induced superselection) of a 
``pointer basis.'' The special role played by (relative) positions in classical physics can be 
traced to the fact that all known interaction Hamiltonians are local (meaning that they 
contain $\delta$~functions of relative positions). Because of this, the spreading of wave 
packets ordained by the uncertainty relations (and speeded up by non-linear dynamics) is 
counterbalanced by an ongoing localization relative to the environment. The result is a 
compromise between localization in position space and localization in momentum space. In 
the appropriate macroscopic limit the hallmark of classicality---localization in phase 
space---emerges. There is every indication that in linear systems these states form an 
overcomplete basis of (near) minimum-uncertainty Gaussian states defining position and 
momentum distributions that afford maximum predictability~\cite{ZurekRG, 
ZurekRevisited, ZurekRevModPhy, BGJKS, JZKGKS, Schlosshauer}. These findings have 
motivated a new interpretation of quantum mechanics, the so-called ``existential 
interpretation''~\cite{ZurekRG, ZurekRevisited, ZurekRevModPhy}, which has even been 
dubbed ``the new orthodoxy''~\cite{Bub}. It is in large part the brainchild of Wojciech H. 
Zurek.

To say that physics is part mathematics and part philosophy is to state the obvious. Without 
the latter, physics would be nothing but a mathematical formalism; any statement 
purporting to address the relation of the mathematical formalism to the physical world is by 
nature philosophical. Philosophy of science comes in two broad categories. Empiricists 
defend the point of view that it is the business of science to make reliable predictions at the 
level of the observable. On this view, quantum mechanics encapsulates correlation laws, and 
the link between the formalism and the real world is measurement outcomes: they are 
(i)~the correlata required by the formalism and (ii)~real. Realists are more ambitious; they 
consider it the aim of science to discover a true description of the world.

In which category does Zurek's interpretation belong? On the one hand, it is clearly more 
ambitious than the correlation interpretation~\cite{Laloe} (a.k.a. ``minimal instrumentalist 
interpretation''~\cite{Redhead}) that satisfies the empiricist philosopher. According to 
Zurek, the existential interpretation completes the most straightforward, most literal, and 
therefore also most simple-minded realist interpretation of quantum mechanics: 
``Decoherence and einselection fit comfortably in the context of the Many Worlds 
Interpretation where they define the `branches' of the universal state vector. Decoherence 
makes MWI complete''~\cite{ZurekRevModPhy}. On the other hand, Zurek appears to want 
to remain noncommittal where the ontological status of quantum states is concerned: 
``There may be in principle a pure state of the Universe including the environment, the 
observer, and the measured system. While this may matter to some~\cite{Zeh}, real 
observers are forced to perceive the Universe the way we do: We are a part of the Universe, 
observing it from within. Hence, for us, {\it environment-induced superselection specifies 
what exists\/}''~\cite[emphasis in original]{ZurekRevModPhy}.

What exists for us, according to Zurek, is predicated on decoherence and einselection, which 
is predicated on the deterministically evolving quantum state associated with a composite 
system that includes the environment and us. What exists for us, therefore, owes its 
existence (for us) to this deterministically evolving quantum state. If, following 
Zeh~\cite{Zeh,ZehBit}, we attribute observer-independent reality to the universal state 
vector, we have an ontological basis on which the ``relative objective existence'' that Zurek 
attributes to pointer states can be predicated. If we were to deny the reality of quantum 
states, we would effectively reduce the quantum formalism to an encapsulation of lawful 
statistical correlations between measurement outcomes. This would not necessarily put us 
inside the empiricist camp, inasmuch as it would allow us to attribute to the correlata an 
absolute objective existence, as the following will show. What makes it impossible to 
reconcile the mathematical formalism of quantum mechanics with the concept of a single 
objective existence is the construal of any one of the theory's formal states as an evolving 
ontological state. One then has two kinds of reality, the bona fide reality of the universal 
state vector, and a reality in scare quotes---``In quantum physics `reality' can be attributed 
to the measured states''~\cite{ZurekRevModPhy}---which belongs to pointer states and to 
measurement outcomes indicated by pointer states. The fact that Zurek is unable to attribute 
to pointer states and measurement outcomes anything stronger than a relative objective 
existence, is to my mind a clear indication of his implicit faith in the absolute reality of the 
universal state vector.

The object of this article is to show how the mathematical formalism of quantum mechanics 
can be reconciled with an unqualified objective existence. Decoherence and einselection are 
as important to making sense of quantum mechanics as Zurek claims they are, but in order 
to arrive at an unqualified objective existence we must free ourselves from three classical 
prejudices: the belief that $\ket{\psi(t)}$ represents a physical state that evolves, the 
belief that $\ket{\psi(t)}$ evolves deterministically, and the belief that space and time are 
intrinsically and infinitely differentiated (Sec.~2). Only by doing so are we in a position to 
descry the ontological implications of the quantum formalism---in particular the relative and 
contingent reality of spatiotemporal distinctions and the extrinsic and finite spatiotemporal 
differentiation of the physical world (Sec.~3)---which are crucial for the wanted 
reconciliation~(Sec.~4).

Contrary to a widespread misconception, viewing the quantum formalism as being 
fundamentally a probability algorithm does not imply that quantum mechanics is concerned 
with states of knowledge rather than states of Nature. On the contrary, unlike 
interpretations that transmogrify the quantal probability algorithm into an evolving physical 
state, it allows us to arrive at a complete and strongly objective description of the physical 
world that requires no reference to observers. What objectively exists, in a sense that 
requires no qualification, is the trajectories of macroscopic objects, whose fuzziness is 
empirically irrelevant, the properties and values of whose possession these trajectories 
provide indelible records, and the fuzzy and temporally undifferentiated states of affairs that 
obtain between measurements and are described by counterfactual probability assignments 
(Sec.~5).

Is it worth the trouble? Is it worth to engage in a struggle against prejudices that are 
hardwired into the neurobiology of perception~\cite{Pseudo, clicks, CCP}, such as the belief 
that space and time are intrinsically and infinitely differentiated? From the empiricist point 
of view, certainly not. For empiricist philosophers, the statement ``there is a teapot on the 
table'' means that anyone who cares to look will see a teapot on the table. The teapot has an 
intersubjective or weakly objective reality, and that's all the reality they require.

Realists want more; they aim at a strongly objective reality independent of observers. If 
they see a teapot on the table then for them there is a teapot on the table {\it whether or 
not\/} anyone cares to look. It has been argued that standard quantum mechanics 
(unadulterated with spontaneous collapses~\cite{GRW, Pearle1, Pearle2} or Bohmian 
trajectories~\cite{Bohm}) is inconsistent with a strongly objective reality, and that we 
must content ourselves with a weakly objective one~\cite{dERP,dEVR,FuPer}. (See, in 
particular, d'Espagnat's assessment of Zurek's philosophy~\cite{dESEP}.) These arguments 
suffer from at least three defects. The first is that nothing but a strongly realist conception 
of the world can explain the miraculous success of ``the pinnacle of human 
thought''~\cite{Zee}, quantum field theory. The second is that science owes its immense 
success in large measure to its powerful ``sustaining myth''~\cite{MerminSM}---the belief 
that we can find out how things {\it really\/} are. Neither the ultraviolet catastrophe nor 
the spectacular failure of Rutherford's model of the atom made physicists question their 
faith in what they can achieve. Instead, Planck and Bohr went on to discover the 
quantization of energy and angular momentum. If today we seem to have reason to question 
our sustaining myth, it ought to be taken as a sign that we are once again making the wrong 
assumptions, and it ought to spur us on to ferret them out. Anything else should be seen for 
what it is---a cop-out. The third defect is that these arguments implicitly accept at least one 
of the three classical prejudices mentioned above.

I make no apologies for the absence of equations in this article. For one thing, the 
quantitative results that underwrite its conclusions are the same as those that underwrite 
the existential interpretation. They are available elsewhere~\cite{ZurekRG, ZurekRevisited, 
ZurekRevModPhy, BGJKS, JZKGKS, Schlosshauer}. For another thing, a physical 
interpretation of the mathematical formalism of quantum mechanics cannot be achieved by 
mathematical means; equations do not address the problem of making physical sense of 
equations.

\section{\large Three classical prejudices}
Every physical interpretation of the mathematical formalism of quantum mechanics has to 
spell out which of the formalism's substructures or structural elements correspond to what 
actually exists. Zurek brings two Significant Insights to bear on this task, which I 
paraphrase by introducing two terms---``extrinsic'' and ``intrinsic''---that seem to me to be 
very useful in elucidating the ontological implications of quantum mechanics.
\begin{enumerate}
\item The values of observables are {\it extrinsic\/}: No value is a possessed value unless 
it is an indelibly recorded value---unless, this is to say, information about an observable's 
value is spread so abundantly across the environment that the resulting decoherence is 
irreversible, at least for all practical purposes.
\item Existence is predicated on persistence. In order to obtain something resembling the 
familiar macroworld, one must attribute objective existence to the maximally predictable 
pointer states. Since these provide their own records (on account of their predictability) they 
can be considered self-existent or {\it intrinsic\/}.
\end{enumerate}
Zurek is compelled to qualify the first insight by turning the possession of a value into 
something that is {\it relative\/} to observers (via the environment in which they are 
embedded), and to qualify the second insight by attributing to the maximally predictable 
pointer states the highest degree of a nevertheless {\it relative\/} objective existence. This 
signals to me that Zurek's insights are marred by three classical prejudices.

The first prejudice is that a vector in Hilbert space represents a state in much the same 
sense that a point in the classical phase space does. One way to surmount this prejudice is 
to look upon both the classical state (a point~$\cP$ in some phase space) and the quantum 
state (a vector~$\ket\psi$ in some Hilbert space) as probability measures. The difference is 
that a classical state assigns trivial probabilities (0~or~1) to the possible values 
(represented by subsets) of every measurable quantity, while a quantum state assigns 
nontrivial probabilities (greater than~0 and less than~1) to the possible values (represented 
by subspaces) of most measurable quantities.

The triviality of the classical probability measure permits its reinterpretation as an objective 
state of affairs: instead of taking $\cP\subset E$ to mean that the probability of finding~$E$ 
is~1, one takes it to imply the truth of ``system~$\cS$ is in possession of property~$E$'' or 
``observable~$O$ has value~$E$''; and instead of taking $\cP\not\subset E$ to mean that 
the probability of finding $E$ is~0, one takes it to imply the falsity of these propositions. If 
one countenances similar inferences for quantum states---$\ket\psi\subset E$ implies the 
truth of these propositions, $\ket\psi\perp E$ implies their falsity---one arrives at the most 
unwanted and unloved aspect of von Neumann's axiomatization of quantum mechanics, the 
collapse postulate for repeatable ideal measurements.

Instead of addressing the root of the disease---the belief that $\ket{\psi(t)}$ represents a 
physical state that evolves---Zurek contents himself with suppressing one symptom of the 
disease, the collapse postulate. This evinces his second classical prejudice, determinism. One 
way to surmount {\it this\/} prejudice is to accept that the phase space formalism of 
classical physics and the Hilbert space formalism of quantum physics both concern lawlike 
correlations between factlike correlata. Here the essential difference is that classical physics 
deals with deterministic correlations while quantum physics deals with 
statistical/probabilistic correlations. From this essential difference further differences 
ensue.

The reason why the classical probability algorithm (represented by a point in some phase 
space) is capable of reinterpretation as an evolving physical state or a changing list of 
possessed properties, is that it assigns only trivial probabilities, and the reason this is so is 
that the classical correlation laws are deterministic: {\it if\/} the state of the system 
is~$\cP_1$ at the time~$t_1$ {\it then\/} it has been or it will be~$\cP_2$ at the 
time~$t_2$. It is worth reminding ourselves that classical physics tells us nothing whatever 
about the state of the system at the time~$t_1$, unconditionally, nor about the mechanism 
or process by which the state at~$t_1$ determines (or is determined by) the state 
at~$t_2$. (Progress in knowledge is often made possible by an admission of ignorance.)

Because the quantal correlation laws are irreducibly probabilistic, a factlike state at~$t_1$ 
is not sufficient to determine a factlike state at~$t_2$. Combined with a factlike state~$\ket 
v$ at~$t_1$ (a set~$v$ of values possessed by a complete set of compatible observables) 
and with the measurement of a complete set of compatible observables at~$t_2$ with 
possible value sets~$w_i$, the relevant correlation law represented by the unitary 
operator~$U(t_2,t_1)$ gives us the probabilities~$|\sandwich{w_k}{U(t_2,t_1)}{v}|^2$ 
with which the factlike state at~$t_2$ turns out to be~$\ket{w_k}$. (As in the classical 
case, nothing needs to be said about the temporal order of $\,t_1$ and~$t_2$.) Without 
assuming that a given set~$W$ of compatible observables is successfully measured at a 
given time~$t_2$, all we have is $|\sandwich{\sqcup}{U(\sqcup,t_1)}{v}|^2$, which has 
two input slots~($\sqcup$), one for a possible value set and one for the time of a 
measurement. This doesn't tell us anything unless we assume the factuality of the value 
set~$v$ at~$t_1$, nor does it tell us anything about what exists, obtains, is factlike, or is 
real at any time other than~$t_1$. (It bears repetition: progress in knowledge is often made 
possible by an admission of ignorance.)

Note that I am not saying anything to the effect that quantum states are states of knowledge 
rather than states of Nature. The fact that the fundamental theoretical framework of physics 
is an irreducible probability algorithm in no wise implies that quantum mechanics is an 
epistemic theory concerned with subjective probabilities. The notion that probabilities are 
inherently subjective is a wholly classical idea. The objective stability of matter (the fact that 
there are stable objects consisting of finite numbers of unextended particles and 
nevertheless occupying finite volumes) does not rest on a subjective uncertainty about the 
position or the momentum of an atomic electron. Quantum mechanics concerns {\it 
objective\/} probabilities (not to be confused with relative frequencies) which have nothing 
to do with how much we know. Subjective probabilities arise if relevant data are ignored; 
they disappear when all relevant data are taken into account. The ``uncertainty'' relations 
guarantee that quantum-mechanical probabilities are objective: they can't be made to 
disappear.

The fact that the fundamental theoretical framework of physics is an algorithm for 
computing objective---and hence fundamentally nonclassical---probabilities, ought to tell us 
something of paramount importance about the physical world, which cannot possibly be 
understood if one treats it so cavalierly as to postulate a fundamentally unobservable, 
unverifiable, and predictively useless determinism. This takes me to Zurek's third classical 
prejudice, which is all but universally shared.

It strikes me as odd that the ontological status of~$\ket\psi$---state of knowledge or state 
of Nature?---has been the focus of a lively controversy for three quarters of a century, while 
the ontological status of the coordinate points and instants on which the wave function 
depends has remained largely unquestioned. (Exceptions are found in the literature on 
quantum gravity, where it is occasionally observed that a fuzzy metric conflicts with the 
postulation of a manifold of points that are sharply localized relative to each other.) If the 
wave function evolves deterministically, it evolves in an intrinsically and infinitely 
differentiated (partitioned) spacetime manifold. If it turned out that the idea of such a 
manifold corresponds to nothing in the physical world, the idea of deterministic evolution 
would share the same fate. Determinism thus implies that space and time 
are---independently of their material ``content''---intrinsically and infinitely differentiated. 
This is what makes it possible to treat them as point sets or as a single manifold. But if this 
is how one conceives of the spatiotemporal aspects of the physical world, one is in no 
position to recognize the ontological implications of the fact that the fundamental theoretical 
framework of physics is an algorithm for computing objective probabilities, as I will show.

If $\ket{\psi}$ represents a physical state, and if it evolves beyond the measurement 
at~$t_2$ without being reset in accordance with the outcome, this outcome (if it exists at 
all) only exists relative to the environment in which it is recorded. For if the quantum state 
of the observed system, the observer, and the environment evolves deterministically and 
contains {\it one\/} possible outcome then it contains {\it all\/} possible outcomes (as a 
superposition). The quantum-mechanical correlation laws ensure that observers who exist in 
the same ``branch'' of the universal state vector and are therefore capable of 
communicating with each other, agree about the outcome. While this saves the classical 
appearances, it obliges Zurek to qualify Significant Insight~\#1 by turning the existence of 
an outcome and the possession of a value into something that is {\it relative\/} to the state 
of the environment and the observers it contains. If predictability is the sole criterion for 
objective existence then all possible outcomes exist, each in a separate branch of the 
universal state vector.

To my way of thinking, treating possible outcomes as actually existing ones is simply a 
category mistake. Possibilities just aren't actualities. I agree with Zurek that predictable 
correlations play a crucial role in the search for those substructures or structural elements of 
the quantum formalism that correspond to what objectively exists: pointer positions have 
values only because their values are predictably correlated (except when they serve to 
record unpredictable values). But predictability of correlations is not sufficient for an 
objective existence that is absolute rather than relative. To arrive at an absolute objective 
existence, one needs to show that objective existence can consistently be attributed to a 
single possible history of the universe. To be able to do this, one needs a criterion for 
deciding when possibilities can ``re-interfere'' and when they cannot. Unless the classical 
prejudices mentioned in this section are eliminated, the possibility of ``re-interference'' 
always exists, at least in principle. But if ``re-interference'' can be ruled out only for all 
practical purposes, no truly indelible record exists, and no value is ever possessed in an 
absolute sense. This obliges Zurek to qualify Significant Insight~\#2 by making objective 
existence quantitatively dependent on the likelihood of ``re-interference'' and thus {\it 
relative\/} in a second, quantitative sense. This likelihood can be quantified by counting the 
number of times information about a value is replicated by the environment, or the number 
of parts of the environment that become correlated with the possible values of an 
observable.

Let us take stock. I have pointed out three classical prejudices and shown how they compel 
Zurek to relativize objective existence in two ways. The first prejudice---the belief that 
$\ket{\psi(t)}$ represents a physical state that evolves---all but implies the second---the 
belief that $\ket{\psi(t)}$ evolves deterministically---for the consequences of the 
alternative---instantaneous ``collapses'' of physical states---are too preposterous for 
consideration, at least in a relativistic world governed by standard quantum mechanics 
(unadulterated with nonlinear modifications of the ``dynamical'' equations~\cite{GRW}). 
The second prejudice entails the third, the belief that space and time are intrinsically and 
infinitely differentiated.

I advocate that theorists think of quantum states the way experimentalists use them, 
namely as algorithms for computing probabilities of possible measurement outcomes on the 
basis of actual measurement outcomes. The quantum ``dynamical'' laws encapsulate 
correlations between measurement outcomes. The expression 
$|\sandwich{\sqcup}{U(\sqcup,t_1)}{v}|^2$ is meaningless unless a possible 
measurement outcome is plugged into the first slot and the time of measurement is plugged 
into the second. I advocate this for several reasons:

(i)~It prevents us from getting mired in disputes over pseudoquestions~\cite{Pseudo}, 
such as how to extract probabilities from the quantum formalism if this is {\it not\/} 
fundamentally a probability algorithm~\cite{probenv}.

(ii)~The axioms of standard quantum mechanics (including the infamous projection 
postulate, stripped of the notion that quantum states are evolving physical states) become 
transparent if the quantum formalism is viewed as an algorithm for calculating objective 
probabilities describing an objective fuzziness such as that on which the stability of matter 
rests~\cite{probenv, justso, IUCAA}. (An algorithm for assigning probabilities to possible 
measurement outcomes on the basis of actual measurement outcomes has two perfectly 
normal dependences. It depends continuously on the times of measurements: if you change 
the time of a measurement by a small amount, the probabilities assigned to the possible 
outcomes change by small amounts. And it depends discontinuously on the results that 
constitute the assignment basis: if you take into account information that was not previously 
taken into account, the assignment basis changes unpredictably as a matter of course.)

(iii)~It becomes possible to recognize ontological implications of the quantum formalism 
that cannot possibly be seen if one starts with metaphysical assumptions that contradict 
them, such as the assumption that the physical world is infinitely differentiated spacewise 
and timewise (Sec.~3).

(iv)~It becomes possible to reconcile quantum mechanics with a single absolute objective 
existence (Sec.~4).

\section{\large The spatiotemporal differentiation of the physical world}
In this section I assume that we can talk unambiguously about objective records, 
measurement outcomes, and possessed values. In so doing I do not take any liberties, for if 
the quantum formalism is fundamentally concerned with correlations between measurement 
outcomes, records, or possessed values then it presupposes the objective existence of these 
things and none of the formalism's ontological implications can compel us to qualify or 
relativize it. One of these ontological implications is that the physical world is {\it not\/} 
infinitely differentiated either spacewise or timewise, as I proceed to explain.

If quantum mechanics is fundamentally an algorithm for computing objective probabilities 
then these probabilities as the formal expression of an objective fuzziness and {\it no object 
ever has a sharp position relative to any other object\/}, except nonrelativistically in the 
unphysical limit of infinite momentum dispersion and infinite mean energy. (The proper 
formalism for dealing quantitatively with fuzzy variables is a probability algorithm. Saying 
that a variable~$Q$ lacks a sharp value is the same as saying that there are finite 
intervals~$I$ for which the proposition ``the value of~$Q$ is in~$I$'' lacks a truth value; 
consequently, no trivial probability can be assigned to the outcome of a measurement of the 
truth value of such a proposition. Note that the meaning of such a proposition is not that 
$Q$ has a precise value somewhere in~$I$ but that $I$ itself is a measured and therefore 
possessed value of~$Q$.)

Consider the probability distribution $|\psi(x)|^2$ associated with the position of the 
electron relative to the nucleus in a stationary state of atomic hydrogen. Imagine a small 
region~$V$ for which $\int_V|\psi(x)|^2dx$ differs from both 0 and~1. While the atom is 
in this state, the electron is neither inside~$V$ nor outside~$V$. (If it were inside, the 
probability of finding it outside would be~0, and if it were outside, the probability of finding 
it inside would be~0.) But being inside and being outside are the only relations that can 
possibly hold between a region~$V$ and an unextended (``pointlike'') object like the 
electron (or the center-of-mass position of an extended object like a C$_{60}$ 
molecule~\cite{Arndtetal}). If neither of these relations holds {\it this region simply does 
not exist for the electron. It has no reality as far as the electron is concerned.\/}

Conceiving of a region~$V$ is tantamount to making the distinction between ``inside~$V$'' 
and ``outside~$V$.'' Hence instead of saying that $V$ does not exist for the electron, we 
may say that the distinction we make between ``inside~$V$'' and ``outside~$V$'' is {\it a 
distinction that the electron does not make.\/} Or we may say that the distinction we make 
between ``the electron is inside~$V$'' and ``the electron is outside~$V$'' is {\it a 
distinction that Nature does not make. It corresponds to nothing in the physical world.} On 
the other hand, if the truth values of these propositions (at a given time~$t$) are measured 
or recorded (in which case one is ``true'' and one is ``false'') one of the possible relations 
between the electron and~$V$ holds at the time~$t$, and the distinction between 
``inside~$V$ at~$t$'' and ``outside~$V$ at~$t$'' is real as far the electron is concerned.

It follows that {\it the reality of spatial distinctions is relative and 
contingent\/}---``relative'' because the distinction we make between the inside and the 
outside of a region may be real for a given object at a given time, and it may have no reality 
for a different object at the same time or for the same object at a different time; and 
``contingent'' because the existence of a given region~$V$ for a given object~$\cO$ at a 
given time~$t$ depends on whether the proposition ``$\cO$~is in~$V$ at the time~$t$'' 
has a (measured or recorded) truth value.

Suppose that $W$ is a region disjoint from~$V$, and that there is a record of $\cO$'s 
presence in~$V$. Isn't $\cO$'s absence from~$W$ implied by this record? Are we not 
entitled to infer that the proposition ``$\cO$~is in~$W$'' has a truth value (namely, 
``false'')? Because the reality of spatial distinctions is relative and contingent, the answer is 
negative. The distinction we make between ``inside~W'' and ``outside~W'' has no physical 
reality {\it per se\/}. If $W$ is not realized (made real) by being the sensitive region of an 
actually existing detector, it isn't available for attribution to~$\cO$, and the same holds for 
the spatial complement $W'$ of~$W$. If neither $W$ nor $W'$ is the sensitive region of an 
actually existing detector, the proposition ``$\cO$~is in~$W$'' cannot have a truth value. 
Therefore all we can infer from $\cO$'s recorded presence in~$V$ is the truth of a {\it 
counterfactual\/}: if~$W$ were the sensitive region of a detector~$D$, $\cO$~would not 
be detected by~$D$.

It follows that a detector---a perfect detector, to be precise, since otherwise the absence of 
a ``click'' does not warrant the falsity of ``$\,O$~is in~$W$''---performs two necessary 
functions: it indicates the truth value of a proposition of the form ``$\cO$~is in~$W$,'' and 
by realizing~$W$ (or the distinction between ``inside~$W$'' and ``outside~$W$'') it 
makes the predicates ``inside~$W$'' and ``outside~$W$'' available for attribution 
to~$\cO$. Much the same applies to spin measurements: the apparatus is needed not only 
to indicate the value of a spin component but also to realize an axis by means of the gradient 
of a magnetic field. The apparatus presupposed by every quantum-mechanical probability 
assignment is needed not only for the purpose of indicating or recording the possession, by 
an observable, of a particular value but also for the purpose of {\it realizing\/} a set of 
values and thereby making them available for attribution. This amply justifies Bohr's 
insistence that, out of relation to experimental arrangements, the properties of quantum 
systems are {\it undefined\/}: positions and orientations need to be possessed in order to 
exist, and they need to exist as properties of measuring equipment in order to be 
attributable as measurement outcomes.

Let \RRR($\cO$) be the set of unpossessed exact positions relative to some object~$\cO$. 
Since no object has a sharp position relative to any other object, we can conceive of a 
partition of \RRR($\cO$) into finite regions that are so small that none of them is the 
sensitive region of an actually existing detector. Hence we can conceive of a partition of 
\RRR($\cO$) into sufficiently small but finite regions~$V_i$ of which the following is true: 
there is no object~$\cQ$ and no region~$V_i$ such that the proposition ``$\cQ$~is 
inside~$V_i$'' has a truth value. In other words, there is no object~$\cQ$ and no 
region~$V_i$ such that $V_i$ exists for~$\cQ$. But a region of space that does not exist for 
{\it any\/} material object, {\it does not exist at all.\/} The regions~$V_i$ represent 
spatial distinctions that Nature does not make. They correspond to nothing in the physical 
world. They exist solely in our heads. Upshot: {\it The physical world is not infinitely 
differentiated spacewise.\/} Its spatial differentiation is finite---it doesn't go ``all the way 
down.''

The same goes for time. The times at which observables possess values, like the possessed 
values themselves, must be recorded in order to exist. Clocks are needed not only to 
measure or record time but also, and in the first place, to make times available for 
attribution to measured or recorded values. Since clocks realize times by the positions of 
their hands, and since exact positions do not exist, neither do exact times. (Digital clocks 
indicate times by transitions from one reading to another, without hands. The uncertainty 
principle for energy and time however implies that these transitions cannot occur at exact 
times, except in the limit of infinite mean energy~\cite{Hilge}.) Exact times are not 
available for attribution. And since the physical world is differentiated timewise by the 
temporal relations (or relative times) that exist in it, its {\it finite temporal 
differentiation\/} follows from the fuzziness of times in exactly the same way that its finite 
spatial differentiation follows from the fuzziness positions.

It follows that physical space cannot be something that exists ``by itself,'' independently of 
its material ``content,'' and that is infinitely differentiated. \RRR~is not an aspect of the 
physical world but an aspect of the quantum-mechanical probability algorithm, useful for 
describing the fuzziness of any existing relative position by a distribution over~\RRR. 
Physical space is something else altogether. It does not ``contain'' matter (which stands to 
reason, since containers have boundaries while space, presumably, has none) but instead is 
an aspect of it. We ought to think of space as the system of spatial relations that hold 
between the world's material constituents (including their relative orientations). There is no 
such thing as empty space, not because space is ``filled with vacuum fluctuations'' but 
because where there is nothing (no thing) there is no {\it there\/}. There are no 
``unoccupied'' positions because there are no unpossessed positions.

How is it that while we readily agree that red, or a smile, cannot exist without a red object 
or a smiling face, we just as readily believe that positions can exist without being properties 
of material objects? We are prepared to think of material objects as substances (things that 
exist ``by themselves,'' without being properties of other things), and we are not prepared 
to think of their properties as substances---except for their positions. The reasons for these 
disparate attitudes are to be found in the neurobiology of perception~\cite{Pseudo, clicks, 
CCP}. They concern the construction of what psychologists and philosophers of mind call the 
``phenomenal world.'' They have nothing to do with the physical world, other than making 
it hard to make sense of it.

The view that all existing positions are possessed relative positions---positions of material 
objects relative to material objects---is well-known and fully consistent with the 
deterministic correlation laws of classical physics~\cite{Dieks1,Dieks2}. Quantum 
mechanics (considered as being fundamentally a probability algorithm) adds to this the 
fuzziness of all existing relative positions, and this makes the relational view of space 
mandatory, as we have just seen. The spatial extension of the physical world, accordingly, is 
not an attribute of a substantial expanse in which spatial relations are embedded but a 
shared attribute of every spatial relation---the quality to which each owes its spatial 
character.

To recap. The fuzziness of all possessed relative positions implies (i)~that the reality of the 
spatial distinctions that we make (equivalent to the objective existence of the regions of 
space that we imagine) is relative and contingent, and (ii)~that the physical world is not 
infinitely differentiated either spacewise or timewise. There is no need to relativize objective 
existence; what is relative---not relative to the components of the universal state vector nor 
relative in a quantitative sense but relative to physical objects---is the reality of the 
distinctions that we make.

The fuzziness of physical observables further implies the extrinsic nature of their values. For 
one thing, if a predicative proposition may or may not have a truth value, a criterion for the 
possession of a truth value is needed. The existence of a (recorded) measurement outcome 
is a sufficient criterion, and in the context of standard quantum mechanics (unadulterated 
with, spontaneous collapses~\cite{GRW,Pearle1,Pearle2}, Bohmian 
trajectories~\cite{Bohm}, or the modal semantical rule~\cite{Dieks3,Dieks4}) it is also the 
necessary criterion. For another thing, if the possible values of positions and orientations do 
not pre-exist as aspects of an intrinsically differentiated space---the same goes for the 
values of the corresponding momenta---then experimental arrangements are needed not 
only to indicate or record values but also to realize them---to make them available for 
attribution to measured systems.

\section{\large One unqualified objective existence}
The existential interpretation postulates deterministic evolution for the state vector or the 
wave function associated with the environment, an observer or observers, and a measured 
system. As we have seen, this entails that the physical world is infinitely differentiated both 
spacewise and timewise, and it compels Zurek to relativize objective existence in two ways. 
The present interpretation (a.k.a. ``the Pondicherry interpretation of quantum 
mechanics''~\cite{WQMITTU, DQSE}) aims at unpacking the presuppositions and 
implications of the quantum formalism, regarded as being fundamentally a probability 
algorithm. One of the presuppositions is the unqualified objective existence of (recorded) 
measurement outcomes---the correlata of the formalism's correlation laws. The nature or 
degree of their objective existence is not an issue. Among the formalism's implications are 
the relative and contingent reality of spatial distinctions and the finite spatiotemporal 
differentiation of the physical world. These implications make it possible to resolve the 
issues peculiar to this approach by adducing the same quantitative results that Zurek 
adduces in support of the existential interpretation, as I proceed to show.

As said, any physical interpretation of the quantum formalism has to identify those 
substructures or structural elements that correspond to what exists. For the empiricist it 
suffices to attribute objective existence to measurement outcomes: everybody knows that 
measurements have outcomes, and quantum mechanics tells us how they are correlated. But 
once the extrinsic nature of the values of observables is taken seriously, a seemingly vicious 
regress arises. No value is a possessed value unless it is an indelibly recorded value, and 
pointer positions are no exception; they have values because their values are inscribed in 
the possessed values of other pointer positions. This seems to send us chasing objective 
existence in never-ending circles, like a dog trying to catch its tail. It is the essence of ``von 
Neumann's catastrophe of infinite regression,'' which has convinced many that it takes an 
extra-physical ontological principle like consciousness to break out of the vicious circle or 
regress~\cite{18errors, LonBau, Wigner, Lockwood, Squires, Albert, Stapp}.

The key to the problem of identifying those substructures or structural elements that 
correspond to what exists, without invoking consciousness, lies in the following facts: while 
all existing positions are fuzzy, some objects have the sharpest positions in existence. This 
has the consequence that the correlations between the recorded positions of most of these 
objects are deterministic in the following sense: the fuzziness of the positions of these 
objects never evinces itself through outcomes or records that are inconsistent with 
predictions that are based on a classical law of motion and earlier outcomes or records.

Let me elaborate. There can be evidence of the departure of an object~$\cO$ from a precise 
trajectory only if there are detectors that can probe the region over which $\cO$'s fuzzy 
position extends. This calls for detectors whose position probability distributions are 
narrower than~$\cO$'s. Such detectors do not exist for all objects. For those objects that 
have the sharpest existing positions, the probability of obtaining a record that is inconsistent 
with a precise trajectory, is necessarily very low. (This follows from the detailed 
investigations mentioned in the Introduction, which underwrite the Pondicherry 
interpretation as well as the existential interpretation.) Hence {\it among\/} those objects 
there are objects of which the following is true: every one of their recorded positions is 
consistent with (i)~every prediction that is based on their previous recorded positions and 
(ii)~a classical law of motion. Such objects deserve to be called {\it macroscopic\/}. To 
enable a macroscopic object to record an unpredictable value, one exception has to be made: 
its position may change unpredictably if and when it serves to record such a value.

Since the positions of macroscopic objects---macroscopic positions, for short---are 
correlated deterministically (except when they serve to record unpredictable values) it is 
possible to pretend that macroscopic objects follow precise trajectories, even during 
value-recording events, when at least one macroscopic position changes unpredictably. In 
reality, though, macroscopic positions are fuzzy like the rest. The difference is that their 
fuzziness never evinces itself through unpredictable records. (Remember that macroscopic 
positions are defined that way.) Macroscopic positions are only counterfactually fuzzy. Their 
fuzziness would evince itself through unpredictable records if the space over which they are 
``smeared out'' were probed. But it never is. This space is undifferentiated; it contains no 
smaller regions. We may imagine smaller regions, but they lack counterparts in the real 
world. The distinctions we make between them are distinctions that Nature does not make. 
Hence when we have said that macroscopic objects follow trajectories that are only 
counterfactually fuzzy, we have said everything that can usefully be said about the positions 
of macroscopic objects. The way we have defined them guarantees that their fuzziness is 
empirically irrelevant.

Now recall that the extrinsic nature of observables is a consequence of their fuzziness, which 
requires a nontrivial probability algorithm for its description. If the fuzziness of macroscopic 
positions is empirically irrelevant then macroscopic positions do not require a nontrivial 
probability algorithm. The triviality of the classical probability measure, as we have seen, 
permits its reinterpretation as an objective state of affairs or a list of possessed properties, 
and so does the triviality of the quantal probability measure when applied to macroscopic 
positions. The reason this is so is not that the probability of finding a macroscopic object 
where classically it could not be, is strictly~0, but that macroscopic objects as we have 
defined them are never found where classically they could not be. The extraordinary 
smallness of said probability, which has been amply demonstrated by decoherence 
investigations, implies that such objects are in plentiful supply: a significant fraction of the 
objects that are commonly called ``macroscopic'' meet our stricter definition. It is therefore 
consistent with the quantal correlation laws to attribute to the (only counterfactually fuzzy) 
values of the positions of macroscopic objects the same objective existence that classical 
physics attributes to the positions of all objects.

Well, almost. While classical positions are intrinsic, in the quantum world even macroscopic 
positions are extrinsic, owing to the mutual dependence of all possessed positions. A 
macroscopic position does not exist ``by itself.'' It is not objective {\it per se\/}. Even the 
Moon has a position only because of the myriad of ``pointer positions'' that betoken its 
whereabouts. Yet the {\it entire\/} system of macroscopic positions is self-contained. There 
is no reason why it should depend on anything external to itself. It is therefore perfectly 
consistent with the quantal correlation laws to attribute to the totality of macroscopic 
positions---the macroworld---an unqualified and independent objective existence.

Given our definition of ``macroscopic object,'' it is clear that we cannot be one hundred 
percent sure that any given object~$\cO$ falls in this category. Even if we had access to 
every existing record of its past whereabouts and knew all relevant boundary conditions, we 
could not completely rule out the possibility of finding it where classically it could not be. 
Even maximally predictable Gaussian states assign small probabilities to such events. It has 
been suggested that such an event would be considered as ``a statistical quirk'' or ``an 
experimental error''~\cite{Peres}. It ought instead to be regarded as confirmation that 
positions are probabilistically correlated. (Such an event must not be confused with the 
unpredictable changes that macroscopic positions undergo when they serve to record 
unpredictable values.)

By our definition, macroscopic positions provide indelible records---not in the sense that the 
theoretical probability of their ``erasure'' is strictly zero but in the sense that they are never 
erased. We therefore have an unambiguous criterion for deciding which possibilities can 
``re-interfere'': in principle all but those that are correlated with the value of a macroscopic 
position. On the other hand, perfect correlation of an observable's possible values with the 
position of what is commonly called a ``macroscopic object'' is not sufficient for the 
existence of an indelible record of the observable's value. If the trajectory of such an object 
can depart unpredictably from the trajectory predicted on the basis of earlier data and a 
classical law of motion then it can also depart unretrodictably from the trajectory retrodicted 
on the basis of later data and a classical law of motion. As a consequence, the trajectory of a 
macroscopic object of common parlance may not always retain information about an 
outcome.

This observation, however, fails to take into account that the positions of macroscopic 
objects of common parlance are abundantly monitored. Suppose that at the time~$t_2$ the 
position of a macroscopic object loses information about an outcome obtained at the earlier 
time~$t_1$. Information about this particular macroscopic position in the interim will be 
retained by the trajectories of a huge number of other macroscopic objects, and among 
these there will be many that conform to our stricter definition of ``macroscopic.'' The 
existence of an indelible record of the outcome of the measurement at~$t_1$ (without 
which there would be no outcome) is therefore not affected by the rare, non-classical 
behavior of what is commonly called a ``macroscopic object.''

To conclude: what objectively exists, in a sense that requires no qualification, is the 
trajectories of macroscopic objects, whose fuzziness is empirically irrelevant, and those 
values of observables of which these trajectories provide indelible records.

\section{\large A quantum system between measurements}
What can we say about the objective properties of quantum systems between successive 
measurements? This question does not arise for a macroscopic object, inasmuch as its 
recorded positions overlap extensively with regard to the times at which they are possessed: 
there is no time span during which no measurement is performed on a macroscopic object.

What about the objective properties of non-macroscopic quantum systems between 
successive measurements? The values of observables being extrinsic, all we can say, and 
{\it all that needs to be said in order to provide an objective description of a quantum 
system between measurements\/}, is what can be inferred from the relevant measurement 
outcomes. What obtains between successive measurements is a fuzzy state of affairs that is 
described---as befits a fuzzy state of affairs---by an algorithm for assigning probabilities to 
the possible results of measurements that might have been but were not performed in the 
meantime. The fact that the quantum formalism is fundamentally a probability algorithm 
thus in no wise implies that it concerns states of knowledge rather than states of Nature, for 
this algorithm describes Nature's objective fuzziness rather than a subjective 
``uncertainty.'' (The literal meaning of Heisenberg's original term `` Unsch\"arfe'' {\it 
is\/} ``fuzziness.'')

The fact that the quantal correlation laws are time-symmetric implies that the fuzzy state of 
affairs that obtains between successive measurements, performed at the respective times 
$t_1$ and $t_2$, is determined not only predictively, by a density operator~$\rho_P(t_1)$ 
that assigns probabilities on the basis of actual outcomes obtained at or before~$t_1$ to the 
possible outcomes of measurements that could have been made after~$t_1$, but also 
retrodictively, by a density operator~$\rho_F(t_2)$ that assigns probabilities on the basis 
of actual outcomes obtained at or after~$t_2$ to the possible outcomes of measurements 
that could have been made before~$t_2$~\cite{DQSE, AhaVaid, RezAha, VaidTSQT}. In 
addition, probability assignments to measurement outcomes at $t>t_1$ made on the basis 
of~$\rho_P(t_1)$ depend on~$U(t,t_1)$, a unitary transformation that encapsulates the 
pertinent diachronic correlations and takes account of the relevant boundary conditions, and 
probabilities assignments to measurement outcomes at~$t<t_2$ made on the basis of 
$\rho_F(t_2)$ depend on~$U(t,t_2)=U^\dagger(t_2,t)$. While probability assignments 
based on earlier {\it or\/} later measurement outcomes are governed by Born's rule, which 
makes use of either $\rho_P$ or~$\rho_F$, probability assignments based on earlier {\it 
and\/} later measurement outcomes are governed by the ABL 
(Aharonov-Bergmann-Lebowitz) rule~\cite{ABL, Kastner}, which makes use of both 
$\rho_P$ and~$\rho_F$.

Unless the relevant Hamiltonian is~0, the probability distributions describing the fuzzy state 
of affairs between successive measurements depend on the times of unperformed 
measurements. This is not the same as saying that the fuzzy state of affairs between actual 
measurements changes with time, for the counterfactual probability assignments describing 
this state of affairs are false not only because they affirm that a measurement is made but 
also because they affirm that this is made {\it at a given time\/}. (Counterfactuals are 
conditional statements whose antecedents are false.) For an unobserved quantum system 
there is no particular time~\cite{DQSE}.

Here our conclusion that the temporal differentiation of the physical world is finite takes on 
a concrete shape. As we have seen, if a region~$V$ is to exist for an object~$\cO$, a 
relation must exist between~$V$ and (the center-of-mass position of)~$\cO$. By the same 
token, if a particular time~$t$ is to exist for~$\cO$, a relation must exist between $t$ 
and~$\cO$. Time, however, is not an observable. It makes no sense to measure the time 
of~$\cO$. What can be measured is the time of possession, by~$\cO$, of a property. 
Accordingly, a particular time $t$~exists for~$\cO$ if and only if it is the (recorded) time of 
possession, by~$\cO$, of a property. The time between successive measurements also 
exists for~$\cO$, but not as a {\it particular\/} time. Instead, it exists {\it as an 
undifferentiated whole\/}. Recall: if the proposition ``$\cO$~is in~$V$ at~$t$'' is true 
while for every $W\subset V$ the proposition ``$\cO$~is in~$W$ at~$t$'' lacks a truth 
value, the distinction we make between~$W$ and its complement in~$V$ has no reality 
for~$\cO$ (at the time~$t$). $V$~exists for~$\cO$ {\it as an undifferentiated whole\/}. 
So does the interval between successive measurements performed on~$\cO$. Where $\cO$ 
is concerned, the state of affairs that obtains between successive measurements is only 
counterfactually differentiated. (It would be differentiated if there were a record of a 
property possessed by~$\cO$ at a particular time in the interim.) $\cO$~is temporally 
differentiated by the measurements that are performed on it.

Obviously, none of these ontological implications of quantum mechanics (qua fundamental 
probability algorithm) can be seen as long as quantum states are regarded as instantaneous 
physical states. An instantaneous physical state implies that the physical world is infinitely 
differentiated timewise, while the quantum-mechanical correlation laws imply that the 
physical world is not infinitely differentiated either spacewise or timewise. In a quantum 
world, there is no such thing as an instantaneous state. Just as the spatial aspect of the 
physical world is not ``build up'' from infinitely small parts (let alone points) so the 
temporal aspect of the physical world is not ``built up'' from instantaneous states.

The conclusion that the state of a quantum system between measurements is temporally 
undifferentiated, and the conclusion that this state is both predictively and retrodictively 
determined, are exceedingly counterintuitive. Our successive experience of the world's 
temporal aspect makes it natural for us to embrace presentism, the view that only the 
present is real, or that it is somehow ``more real'' than the future or the past. Our 
self-experience as agents makes it natural for us to hold that the known or in principle 
knowable past is ``fixed and settled,'' while the unknown and apparently unknowable future 
is ``open.'' None of this has anything to do with the physical world---the world as described 
by physics. It is impossible to consistently project the sole or greater reality of the 
experiential now into the objective physical world. To philosophers, the perplexities and 
absurdities entailed by the notion of a changing objective present are well known. (See, e.g., 
the illuminating entry on ``time'' in Ref.~\cite{Audi}.) To physicists, the subjectivity of a 
temporally unextended yet persistent and persistently changing present was brought home 
by the relativity of simultaneity. Much the same is implied by quantum mechanics, inasmuch 
as this tells us that the temporal aspect of the physical world cannot be built up from 
instantaneous states.

Nor does physics know anything about a preferred direction of causality. It is we who base 
on our sense of agency, our ability to know the past, and our inability to know the future, the 
figment of a causal arrow. The physical correlation laws are time-symmetric. They let us 
retrodict as well as predict. The fact that the undifferentiated and fuzzy state of affairs that 
obtains between measurements is determined retrodictively as well as predictively becomes 
incomprehensible only if we combine the figment of a causal arrow with the figment of an 
instantaneous state and project the result---an evolving instantaneous state---into the 
physical world. This leads to the well-known folk tale according to which causal influences 
reach from the nonexistent past to the nonexistent future through persisting ``imprints'' on 
the present. If the past and the future are unreal, the past can influence the future only 
through the mediation of something that persists. Causal influences reach from the past into 
the future by being ``carried through time'' by something that ``stays in the present.'' This 
evolving instantaneous state includes not only all presently possessed properties but also 
traces of everything in the past that is causally relevant to the future. In classical physics 
this is how we come to conceive of fields of force that evolve in time (and therefore, in a 
relativistic world, according to the principle of local action), and that mediate between the 
past and the future (and therefore, in a relativistic world, between local causes and their 
distant effects). In quantum physics, this is how we come to seize on a probability algorithm 
that depends on the time of a measurement and the results of earlier measurements and to 
transmogrify the same into an evolving instantaneous state.

\section{\large Summary}
Zurek's interpretation of quantum mechanics suffers from three classical prejudices, which 
necessitate a twofold relativization of the concept of objective existence: the belief that 
$\ket{\psi(t)}$ represents a physical state that evolves, the belief that $\ket{\psi(t)}$ 
evolves deterministically, and the belief that space and time are intrinsically and infinitely 
differentiated. If instead the quantum formalism is viewed as an algorithm for calculating 
objective probabilities, describing an objective fuzziness, we are in a position to recognize 
its ontological implications---in particular the relative and contingent reality of 
spatiotemporal distinctions and the extrinsic and finite spatiotemporal differentiation of the 
physical world. This makes it possible to reconcile quantum mechanics with a single absolute 
objective existence. Contrary to a widespread misconception, viewing the quantum 
formalism as being fundamentally a probability algorithm does not imply that quantum 
mechanics is concerned with states of knowledge rather than states of Nature. On the 
contrary, it makes possible a complete and strongly objective description of the physical 
world that requires no reference to observers. What objectively exists, in a sense that 
requires no qualification, is the trajectories of macroscopic objects, whose fuzziness is 
empirically irrelevant, the properties and values of whose possession these trajectories 
provide indelible records, and the fuzzy and temporally undifferentiated states of affairs that 
obtain between measurements and are described by counterfactual probability assignments. 
The quantitative results that underwrite these conclusions are the same as those that 
underwrite the existential interpretation.

\end{document}